\documentclass[english,aps,prl,twocolumn,showpacs,preprintnumbers,amsmath,amssymb,showkeys]{revtex4}

\usepackage[T1]{fontenc}
\usepackage{color}
\usepackage{units}
\usepackage{amssymb}
\usepackage{graphicx}
\usepackage{esint}
\usepackage{bm}
\usepackage{natbib}
\usepackage{ulem}
\usepackage{amsmath}
\usepackage[colorlinks=true, citecolor=red]{hyperref}
\setcitestyle{journalcolor= blue}

\usepackage{graphicx}% Include figure files
\usepackage{dcolumn}% Align table columns on decimal point
\usepackage{bm}% bold math

\usepackage{babel}

\begin{document}
	
	\title{Anharmonicity in Raman active phonon modes of atomically-thin MoS$_2$}
	
    \author{Suman Sarkar}
    %\thanks{These two authors contributed equally}
    \author{Indrajit Maity}
    %\thanks{These two authors contributed equally}
    \author{H.L. Pradeepa}
    \affiliation{Department of Physics, Indian Institute of Science, Bangalore-560012, India}
	\author{Goutham Nayak}
	\author{Laetitia Marty}
	\author{Julien Renard}
	\author{Johann Coraux}
	\author{Nedjma Bendiab}
	\author{Vincent Bouchiat}
	\affiliation{Univ. Grenoble Alpes, CNRS, Grenoble INP, Institut N\'eel, 38000 Grenoble, France}
	\author{Sarthak Das}
	\author{Kausik Majumdar}
	\affiliation{Department of Electrical Communication Engineering, Indian Institute of Science, Bangalore 560012, India}
	
	\author{Manish Jain}
		\author{Aveek Bid}
	\email{aveek@iisc.ac.in}
	\affiliation{Department of Physics, Indian Institute of Science, Bangalore 560012, India}

	\begin{abstract}
Phonon-phonon anharmonic effects have a strong influence on the phonon spectrum; most prominent manifestation of these effects are the softening (shift in frequency) and broadening (change in FWHM) of the phonon modes at finite temperature. Using Raman spectroscopy, we studied the temperature dependence of the FWHM and Raman shift of $\mathrm{E_{2g}^1}$ and $\mathrm{A_{1g}}$ modes for single-layer and natural bilayer MoS$ _{2} $ over a broad range of temperatures ($8 < $T$ < 300$ K). Both the Raman shift and FWHM of these modes show linear temperature dependence for $T>100$~K, whereas they become independent of temperature for $T<100$~K. Using first-principles calculations, we show that three-phonon anharmonic effects intrinsic to the material can account for the observed temperature-dependence of the line-width of both the modes. It also plays an important role in determining the temperature-dependence of the frequency of the Raman modes. The observed evolution of the line-width of the A$_{1g}$ mode suggests that electron-phonon processes are additionally involved. From the analysis of the temperature-dependent Raman spectra of MoS$_2$ on two different substrates -- SiO$_2$ and hexagonal boron nitride, we disentangle the contributions of external stress and internal impurities to these phonon-related processes.  We find that the renormalization of the phonon mode frequencies on different substrates is governed by strain and intrinsic doping. Our work establishes the role of intrinsic phonon anharmonic effects in deciding the Raman shift in MoS$ _{2} $ irrespective of substrate and layer number.

	\end{abstract}
	\maketitle

\section{Introduction}
MoS$_2$ is a well-studied two-dimensional transition metal dichalcogenide having a direct bandgap in its single layer form~\cite{mak2010atomically}. Its discovery~\cite{mak2010atomically,radisavljevic2011single} has opened up new possibilities for the semiconductor industry in terms of next-generation optoelectronics and valleytronics devices. The predicted, and observed exotic properties  of atomically-thin flakes of MoS$_2$ -- direct bandgap in the monolayer limit, strong correlations  giving rise to three-body trion state observable even at room-temperature~\cite{mak2013tightly}, ambipolar transport~\cite{bao2013high,zhang2012ambipolar},  superconductivity~\cite{lu2015evidence,taniguchi2012electric}, highly efficient  light-matter interactions~\cite{article} and strong valley-selectivity~\cite{Mak2012,Xu2014} -- make it a very interesting  material from an academic perspective. Optical measurements like Raman and photoluminescence (PL) spectroscopy have become the techniques of choice to probe these properties of MoS$_2$ and other related transition metal dichalcogenides. 

These optical properties of these materials are influenced by presence of intrinsic defects (primarily chalcogenide vacancies)~\cite{hong2015exploring,tongay2013defects,ong2013mobility},  effect of ambient, the substrate or capping layer~\cite{doi:10.1021/acsnano.7b05520,tongay2013defects}, temperature, ~\cite{molina2013effect,latzke2015electronic}. Analyzing the optical properties thus provide a wealth of information about defect-dynamics,  various energy transfer processes, Coulomb interactions, influence of electric, magnetic field and strain on spin- and valley-splitting of the electronic energy bands~\cite{RevModPhys.90.021001}  and temperature-dependent multi-phonon scattering processes~\cite{golasa2014multiphonon}. Additionally, valuable information can be gleaned from the peak-positions and full width at half-maximum (FWHM) of the Raman-active modes about the dielectric environment~\cite{lin2014dielectric,yan2012raman}, strain effects ~\cite{rice2013raman,wang2013raman,castellanos2013local}, anharmonicity in the lattice potential energy~\cite{boukhicha2013anharmonic,yang2017anharmonicity,sahoo2013temperature}, thermal expansion~\cite{huang2014correlation,late2014thermal,sahoo2013temperature}, and thermal conductivity~\cite{liu2013phonon,li2013thermal,sahoo2013temperature,thripuranthaka2014temperature} of these materials.  Characterizing the effect of the latter (temperature) is essential to understand the limitations to charge carrier mobility and thermal conductivity, hence the performance of devices based on these materials. To this respect, interactions involving two or more phonons ~\cite{golasa2014multiphonon} and electron-phonon interactions are essential. Their characteristic temperature-dependence allows to unravel their role.    
	
Even after intense research for more than a decade, an in-depth, combined experimental and theoretical study of the temperature evolution of the phonon-modes; in single- and few-layer MoS$_2$ is missing. Previous studies employed semi-quantitative models for the calculation of the temperature dependent Raman shift \cite{najmaei2013quantitative,yang2016excitation}. Two contradictory conclusions have been reported by fitting the experimental data to such models for 300~K$<T<$500~K.  Najmei et-al. concluded that the three-phonon process are irrelevant in explaining the observed phonon-frequency shifts of Raman active modes, whereas Yang et-al.  found the three-phonon processes to be more dominant \cite{najmaei2013quantitative,yang2016excitation}. Lanzillo et-al. \cite{Lanzillo_apl_2013} performed first-principles molecular dynamics simulations  including both the three- and four-phonon processes, predicting a characteristic temperature-dependence of the Raman shift. However testing this scenario against experimental data was made difficult by the rather limited resolution ($\sim$2~cm$^{-1}$) of the measured Raman spectra. Studies conducted at high temperatures (300~K<$T$<450~K) show that the  positions of the Raman peaks of single-layer MoS$_2$ are linearly red-shifted with increasing temperature~\cite{doi:10.1021/acsami.5b00690, doi:10.1021/nn405826k,doi:10.1021/jp402509w,doi:10.1063/1.4819337}. This shift was attributed primarily to four-phonon processes for the $\mathrm{E_{2g}^1}$ mode while for the $\mathrm{A_{1g}}$ mode, thermal expansion was also found to play a significant role~\cite{C3NR02567E}. 

In this letter, we report the results of combined experimental and theoretical studies of the temperature dependence of the Raman shifts and full-width at half maxima (FWHM) of Raman-active modes in single-layer (SL) and natural bilayer (NBL)  MoS$_2$ samples. We find that at low temperatures, the FWHM and Raman shift are practically independent of $T$ while above 125~K they vary almost linearly with temperature.  
Our theoretical calculations, based on density functional perturbation theory~\cite{baroni2001phonons} (DFPT), faithfully reproduce the observed temperature dependencies of FWHM of the Raman active vibrational modes of SL MoS$_2$ over the entire temperature range. We find that three-phonon anharmonic effects are the predominant factor determining the observed temperature dependence of the FWHM for the two prominent Raman modes in MoS$_2$. We also find that  higher order phonon processes contribute significantly to the observed high-temperature softening of phonon modes with the temperature. By comparing the temperature dependence of the FWHM and the Raman shift  samples prepared on two types of substrate --  SiO$_2$/Si$^{++}$  and single crystal hexagonal boron nitride (hBN) -- we find that  strain and electronic doping only contribute a temperature-independent change of the Raman-shifts and FWHMs.
	
\section{Methods and Measurements}
	\begin{figure}[h]
		\begin{center}
			\includegraphics[width=0.5\textwidth]{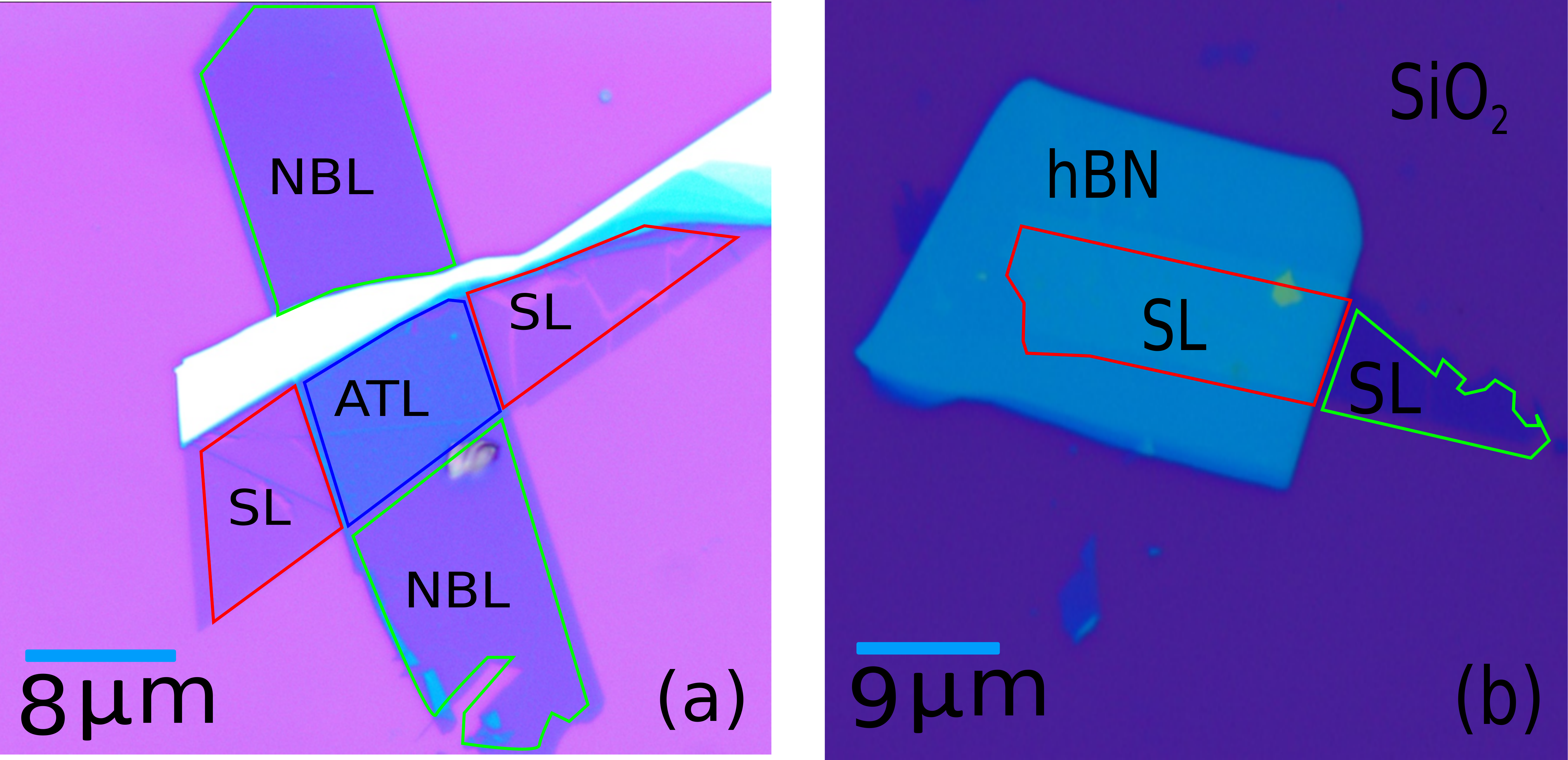}
			\small{{\caption{(a) Optical image of Sample1 prepared on SiO$_2$/Si$^{++}$ substrate -- the  SL region is outlined in solid-red line and NBL in solid-green line. (b) Optical image of Sample2 prepared on a hBN sample -- the MoS$_2$ on hBN portion is outlined by a solid-red line and the MoS$_2$ on SiO$_2$ by a solid-green line. \label{Fig:1}}}}
		\end{center}	
	\end{figure} 

 MoS$_2$ flakes were exfoliated from single crystals of naturally occurring  MoS$_2$ (SPI supplies) on polydimethylsiloxane (PDMS). These were transferred on 290~nm SiO$_2$/Si$^{++}$ substrates using the well-established dry-transfer technique~\cite{2053-1583-1-1-011002, dean2010boron}.   We also prepared hBN/MoS$_2$ heterostructures on SiO$_2$/Si$^{++}$ substrates by sequentially aligning and transferring the different atomic layers.  The transfers were carried out in an inert atmosphere inside a glove-box using a custom-built system based on a Thorlabs  3-Axis  motorized linear-translation stage (MTS-Z8) under an optical microscope. During transfers, the speed of approach and retraction of the microscope stage was kept less than 1~$\mu $m s$^{-1}$ to avoid wrinkling and tearing of the flakes. The number of layers in the flakes was initially estimated from the colour-contrast of the optical images and later confirmed from AFM measurements, Raman spectroscopy, and photoluminescence measurements. We have presented high-frequency and low-frequency Raman measurement along with PL measurement on appendix section Fig~\ref{Fig:7} to identify the number of layers in our sample. We also have identified SL and NBL sample with low-frequency Raman measurement as discussed in the appendix section in the Fig~\ref{Fig:8}.

We studied two different classes of samples. Multiple samples of each type were studied, and the data on different samples of the same class were qualitatively consistent. In this letter, we focus on the data obtained from a single sample of each class. Fig~\ref{Fig:1}(a) is an optical image of a sample prepared on a SiO$_2$/Si$^{++}$ substrate  having SL, NBL and artificial trilayer (ATL) regions (henceforth referred to as `Sample1'). On this class of samples, we studied the temperature evolution of Raman spectra for the SL and NBL. In the second class of samples, SL MoS$_2$ was transferred to lie partially on SiO$_2$  and partly on a $\sim$20~nm thick hBN flake on the surface of SiO$_2$. An optical image of one such sample (referred to as `Sample2')  is shown in Fig.~\ref{Fig:1}(b). These samples were used for the comparative study of the temperature dependence of Raman spectra of SL MoS$_2$ on SiO$_2$ and hBN.

We have performed the low-temperature Raman experiments in reflectance mode on HORIBA Scientific Instrument LABRAM HR Evolution attached with a ARS cryo-free cryostat which can reach till 4K base temperature. 532~nm laser source in our measurement system having a spatial resolution of 1~$\mu$m is been used. The Raman spectra were recorded using 1800 lines/mm grating at very low laser power levels ($\sim$~100~ $\mu$W) to avoid heating of the sample. The resolution of our set up is about 1 cm$^{-1}$. In this report we have concentrated our result over the spectral-range 350-450~cm$^{-1}$ though we have measured the spectra over the range of spectral-range 200-800~cm$^{-1}$. In the appendix section we have shown in Fig.~\ref{Fig:6} a typical Raman spectra over complete range taken at 300~K.

The  phonon frequencies, Raman shifts and FWHM were computed using DFPT~\cite{baroni2001phonons} and the D3Q code~\cite{paulatto2013anharmonic} based on density functional theory as implemented in the QUANTUM ESPRESSO package~\cite{Kohn_dft_1965, giannozzi2017advanced,giannozzi2009quantum}. We used optimized norm-conserving Vanderbilt (ONCV) pseudopotentials~\cite{hamann2013optimized,schlipf2015optimization} for Mo and S atoms and the generalized gradient approximation~\cite{perdew1996generalized} for the exchange-correlation. For $\mathrm{SL \ MoS_{2}}$ and $\mathrm{NBL \ MoS_{2}}$, we used plane-waves kinetic energy cut-off of 80 Ry ensuring convergence of phonon frequencies at $\Gamma$ point within 0.1 $\mathrm{cm^{-1}}$. To avoid interaction between periodic images along $z$-direction, an 18 {\AA} vacuum spacing was used. For total energy calculations, we used a $12\times12\times1 \ \vec{k}$ point sampling of the Brillouin zone, whereas the phonon frequencies, third order force constants were calculated using a $3\times3\times1 \ \vec{k}$ point sampling of the Brillouin zone. Raman shift and FWHM calculations using the third order force constants were performed by summing over  discrete uniform $\vec{q}$ grid points ($200\times200\times1$) which were randomly shifted from the origin. A Gaussian function with a smearing of $3 \ \mathrm{cm^{-1}}$ was used \cite{paulatto2013anharmonic} to replace the delta function.
	 	
\begin{figure}[t]
	\begin{center}
		\includegraphics[width=0.5\textwidth]{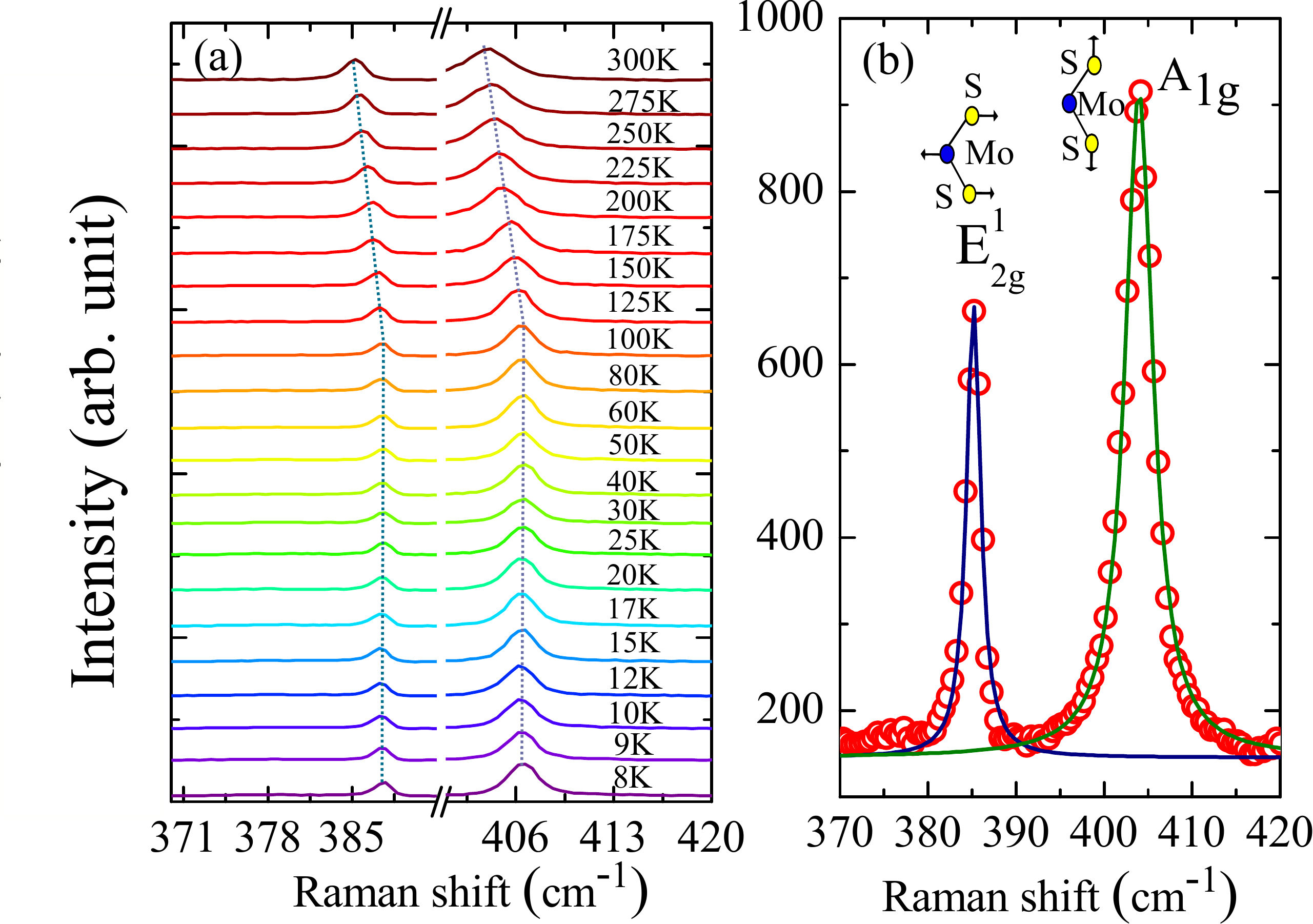}
		\small{{\caption{(a). Raman spectra of Sample1 measured at different temperatures between 8~K and 300~K. The dotted lines are the guide for the eye for marking the evolution of Raman shift with temperature. (b) An individual Raman spectra at 300~K. The blue and olive solid lines are Lorentzian fits to the experimental data. The inset schematics show the $\mathrm{E_{2g}^1}$ and $\mathrm{A_{1g}}$ vibrational modes.     \label{Fig:2}}}}
	\end{center}
\end{figure} 
\section{Results and discussions}
\begin{figure}[t]
	\begin{center}
		\includegraphics[width=0.5\textwidth]{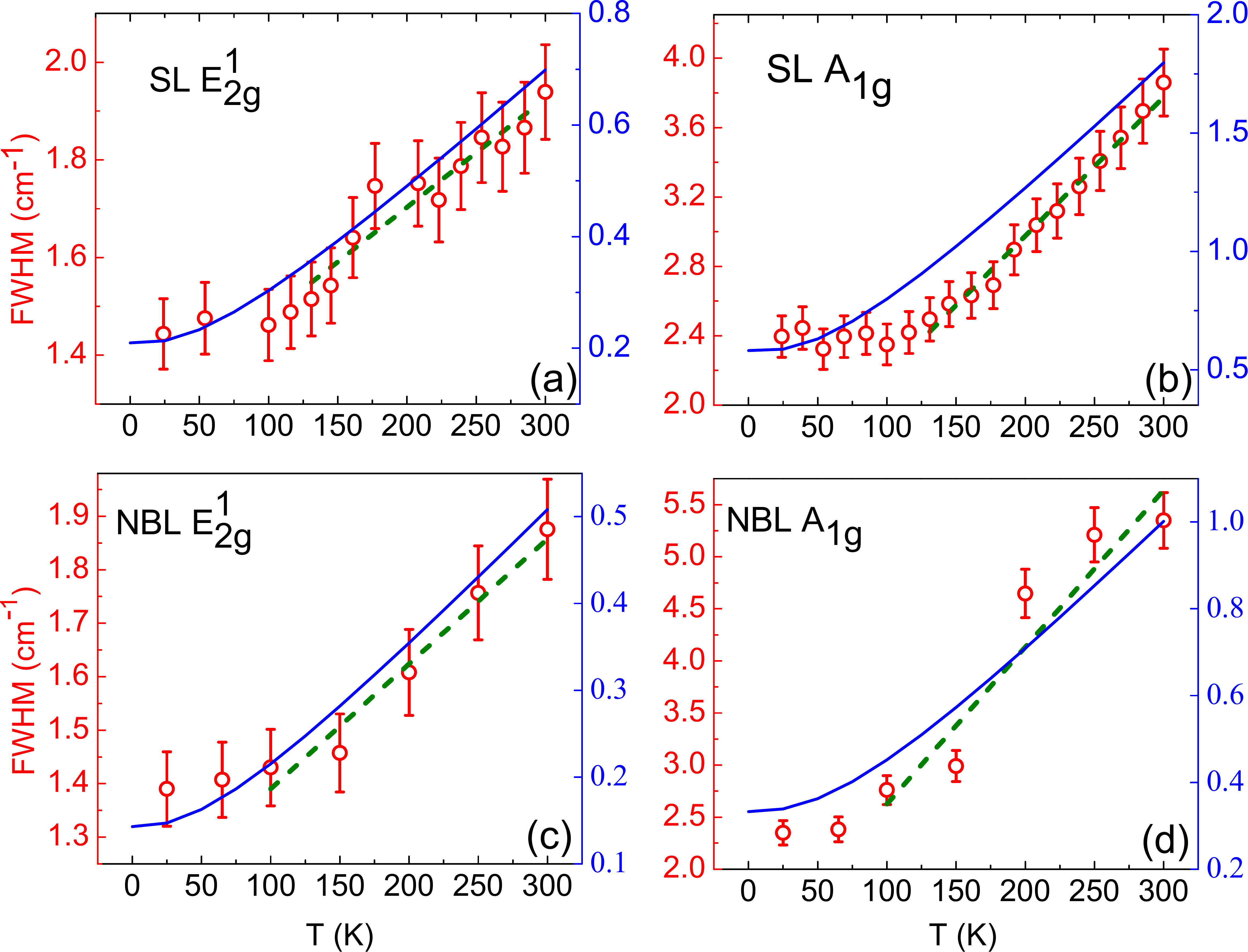}
		\small{{\caption{Temperature dependence of the FWHM of the Raman modes. The experimentally obtained data are plotted on the left-axis in open red circles for (a) the $\mathrm{E_{2g}^1}$ peak of the SL region, (b) the $\mathrm{A_{1g}}$ peak of  the SL region. The corresponding theoretical results are plotted on the right-axis  in solid blue lines. The green dashed lines are  linear fits to the experimental data for $T>100$~K. (c), (d) are the corresponding plot for NBL region. The data are for Sample1.
					\label{Fig:3}}}}
	\end{center}
\end{figure}

\subsection{Temperature dependence of the life-times of Raman modes in MoS$_2$}	
 High frequency Raman spectrum (we define high-frequency in this letter to be the spectral-range 350-450~cm$^{-1}$) of SL and NBL MoS$_2$ consists of two Raman active modes -- denoted by $E'$ and $A'_{1}$ for SL and by $\mathrm{E_{2g}^1}$ and $\mathrm{A_{1g}}$ for multi-layers~\cite{PhysRevB.84.155413, doi:10.1002/adfm.201102111,saito2016raman, doi:10.1021/nn1003937}. The E modes arises from in-plane, anti-phase oscillations of the two $\mathrm{S}$ atoms with respect to the $\mathrm{Mo}$ atom while the $\mathrm{A}$  modes are due to the anti-phase, out-of-plane oscillations of only the $\mathrm{S}$ atoms~\cite{li2012bulk}. For notational simplicity,  we will refer to the in-plane  modes as $\mathrm{E_{2g}^1}$ and the out-of-plane modes as $\mathrm{A_{1g}}$ for both SL and multi-layers. The measured Raman spectra in the range 370 - 420 $\mathrm{cm^{-1}}$ over the temperature range 8~K to 300~K are shown in Fig.~\ref{Fig:2}(a) for SL region for Sample1.  The dotted lines are guides to the eye showing the evolution of the peak positions with temperature $T$. We found that even in the SL limit, our $\mathrm{MoS_{2}}$ samples have FWHM of about $\mathrm{ \sim 1.4 \ cm^{-1} }$ and $\mathrm{ \sim 2.3 \ cm^{-1} }$ for $\mathrm{E_{2g}^1}$ and $\mathrm{A_{1g}} $ respectively which confirms  the high crystalline quality of the $\mathrm{MoS_{2}}$ flakes~\cite{doi:10.1021/jp402509w}.

To extract the peak positions and peak FWHM, we fitted the Raman spectra at every temperature with two Lorentzian peaks. An example is shown in Fig.~\ref{Fig:2}(b) for the data obtained at 300~K. We have also used Voigt fitting to determine the peak position and FWHM shown in appendix section Fig.~\ref{Fig:9} and Fig.~\ref{Fig:10} . We found that in case of Voigt fitting both Lorentzian width and Gaussian width contribute to the Voigt width and as a result we end up with a minimum FWHM for $\mathrm{E_{2g}^1}$ and $\mathrm{A_{1g}} $  peak is about $\mathrm{ \sim 1.8 \ cm^{-1} }$ and $\mathrm{ \sim 2.5 \ cm^{-1} }$ respectively while from Lorentzian fit to our Raman spectra provides the minimum FWHM is about $\mathrm{ \sim 1.45 \ cm^{-1} }$ and $\mathrm{ \sim 2.3 \ cm^{-1} }$ respectively shown Fig.~\ref{Fig:10} (a) and (b) respectively.  In Fig.~\ref{Fig:3}(a) and (b) we show the temperature dependence of the FWHM of the $\mathrm{E_{2g}^1}$ and $\mathrm{A_{1g}}$ mode respectively for the SL region of Sample1. We observe that the FWHM decreases with decreasing $T$ and tends to saturate below 100~K. The same trend is seen for the FWHM of the $\mathrm{E_{2g}^1}$ and $\mathrm{A_{1g}}$ modes measured on the NBL portion of Sample1 -- the data are plotted in Figs.~\ref{Fig:3}(c) and (d) respectively.

Raman modes have a finite FWHM due to both intrinsic phonon scattering processes  and extrinsic factors (linked to defects, the finite size of crystals, etc). At a finite temperature, the intrinsic FWHM of the Raman modes $\gamma^{in}$  are determined by both electron-phonon interactions and phonon-phonon anharmonic effects i.e. $\gamma^{in}=\gamma^{e-ph}+\gamma^{ph-ph}$~\cite{su2014dependence,sahoo2013temperature}. Extrinsic factors linked to defect-dynamics can cause broadening of the Raman lines in several possible ways --  by changing the anharmonicity,  causing local fluctuations of the frequency of optical-phonons,  causing phonon confinement-induced relaxation of the Raman wave vector selection rules or leading to a change of phonon wave-function as the solution of the dynamic problem~\cite{cryst7080239}.  

We compute the FWHM using third-order phonon-phonon anharmonic effects for both SL and NBL samples (the details of the calculations are given in the following section).
The calculated change in FWHM over the temperature range $0<T<300$~K of $\mathrm{SL \ \ MoS_{2}}$ ($\mathrm{\sim 0.48\  cm^{-1} and \ 1.2 \ cm^{-1}}$ and for $\mathrm{E_{2g}^1}$ and  $\mathrm{A_{1g}}$ respectively) matches quite well with our measured values ($\mathrm{\sim 0.5\  cm^{-1} and \ 1.55 \ cm^{-1}}$ for $\mathrm{E_{2g}^1}$ and  $\mathrm{A_{1g}}$ respectively). The data are plotted in Figs.~\ref{Fig:3}(a) and \ref{Fig:3}(b) respectively for the two modes.  

Just like the experimental data the calculated temperature dependence of FWHM of both $\mathrm{A_{1g}}$ and $\mathrm{E_{2g}^1}$ modes for SL $\mathrm{MoS_{2}}$, saturates below 100~K and increases linearly for  $T>100$~K with a coefficient we call $\beta$. The calculated and measured values of $\beta$ of the $\mathrm{E_{2g}^1}$ and $\mathrm{A_{1g}}$ peaks for SL and NBL are presented in  Table.~\ref{tab:1} while the calculated value of $\beta$ is in good agreement with experiment for in-plane vibrational mode, it is underestimated for the out-of-plane mode. This discrepancy can not be assigned to four-phonon or higher order phonon processes. If four-phonon or higher order phonon processes played an important role then both the modes should have shown the discrepancy. In contrast, the electron-phonon process mostly affects the $\mathrm{A_{1g}}$ modes by reducing its life-time, hence by increasing the FWHM. Naturally, our calculations addressing the case of undoped MoS$_2$ cannot properly account for the observed electron-phonon interaction-related increase of the FWHM.

\subsection{Temperature dependence of frequency of Raman modes in MoS$_2$}

\begin{figure}[t]
	\begin{center}
		\includegraphics[width=0.5\textwidth]{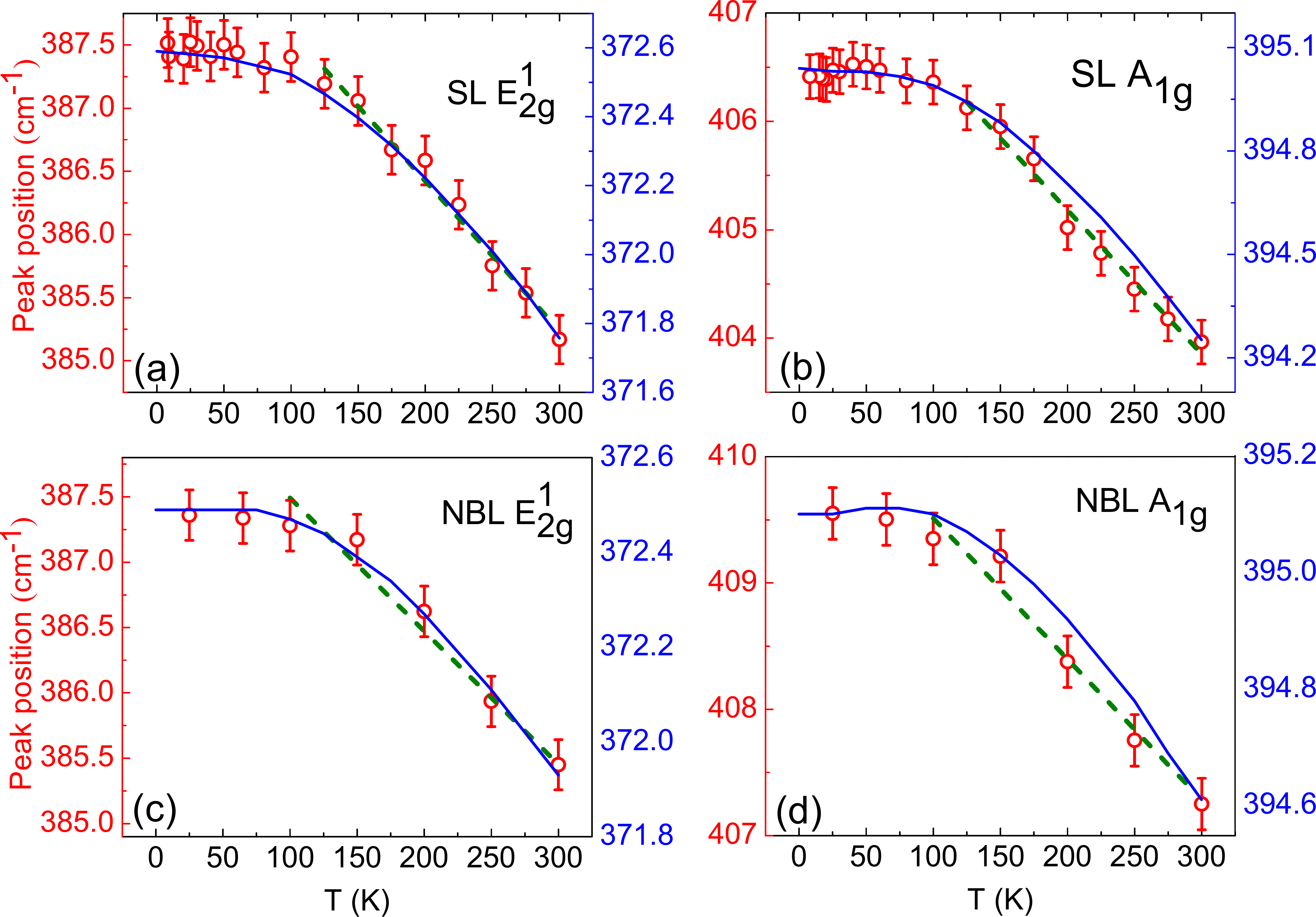}
		\small{{\caption{Temperature dependence of the Raman shift. The experimentally obtained data are plotted on the left-axis in open red circles for (a) the $\mathrm{E_{2g}^1}$ peak of the SL region,  (b) the $\mathrm{A_{1g}}$ peak of the SL region, (c) the $\mathrm{E_{2g}^1}$ peak of the NBL region, and (d)  the $\mathrm{A_{1g}}$ region of the NBL sample. The corresponding theoretical results are plotted on the right-axis  in solid blue lines. The measurements were performed on Sample1. The green dashed lines are the linear fits to experimental data for $T>100$~K.  \label{Fig:4}}}}
	\end{center}
\end{figure}

The experimentally measured temperature dependence of the Raman shifts of the $\mathrm{E_{2g}^1}$ and  $\mathrm{A_{1g}}$ modes from 300~K down to 10~K in the SL region of Sample1 are shown in Figs.~\ref{Fig:4}(a) and (b) respectively. The corresponding data for the NBL region of Sample1 are shown in Figs.~\ref{Fig:4}(c) and (d), respectively. We find that above 100~K, the Raman shift of both the $\mathrm{E_{2g}^1}$ and  $\mathrm{A_{1g}}$ modes, in both SL and NBL, decrease with temperature. Below 100~K, the Raman shift in all cases saturates. Over the temperature range from 100~K to 300~K the Raman shifts can be fitted to a relation:

\begin{eqnarray}
\omega(T)=\omega_0 - \alpha T
\end{eqnarray}
Here $\omega_0$ is the peak-position extrapolated to zero temperature and $\alpha = d\omega/dT$ is the temperature coefficient of the Raman shift. From linear fits to the data for $T>100$~K, we obtain, for both SL and NBL, $\alpha -\mathrm{\sim 0.013\ cm^{-1}/K}$ for the $\mathrm{E_{2g}^1}$ mode and $\mathrm{\sim 0.012\ cm^{-1}/K}$ for the $\mathrm{A_{1g}}$ mode. In the appendix section, we have shown that the Voigt fitting to the Raman spectra does not gave any appreciable difference to the peak-position shown in Fig.~\ref{Fig:9} (a) and (b).
 
To understand the temperature dependence of the Raman shifts, note that at $T\neq 0$ K, phonons frequencies get renormalized due to anharmonic effects. At constant (external) pressure, the corresponding temperature dependence of the phonon frequencies $\omega$ for a semiconductor can be expressed as $\omega(T) - \omega(0) = \Delta \omega_{T} (T) + \Delta \omega_{V} (T)$, where $\Delta \omega_{T}$ and $\Delta \omega_{V}$ are the frequency shifts due to lattice thermal expansion and `pure' thermal anharmonic effects, which typically include third- and fourth-order phonon-phonon  anharmonic effects \cite{balkanski1983anharmonic, menendez1984temperature,klemens1966anharmonic}. We compute the frequency shifts by including both the lattice thermal expansion effects~\cite{najmaei2013quantitative} and phonon-phonon anharmonic effects, using three-phonon processes. Two kinds of scattering processes are taken into account: (a) a phonon of momentum $\vec{q}$  can decay into two phonons ($-\vec{q^{'}}$, $-\vec{q^{''}}$), (b) a phonon with $\vec{q}$  can coalesce with a phonon with $-\vec{q^{'}}$ to eject one with $-\vec{q^{''}}$. In our first principles based calculations we  incorporated all such possible three-phonon processes without any restriction~\cite{paulatto2013anharmonic,menendez1984temperature}. 

In Figs.~\ref{Fig:4} (a) and (b), we plot, on the right-axis, the calculated temperature dependence of the $\mathrm{E^{1}_{2g}}$ and $\mathrm{A_{1g}}$ Raman shifts for SL $\mathrm{MoS_{2}}$. For $T\gtrsim100$ K, the Raman modes soften linearly with $T$ while for $T\lesssim 100$ K, the Raman shifts saturate and becomes independent of $T$. This theoretically computed trend  captures correctly the experimental temperature dependence.

The temperature dependencies of Raman shifts and FWHM arise from the occupation-probabilities of the phonons (Bose factor)~\cite{menendez1984temperature,paulatto2013anharmonic}. In order to get an intuitive  understanding of the temperature dependence of the Raman shift, let us consider one possible three-phonon decay channel: an optical phonon with energy $\hbar\omega_{0}$ at $\Gamma$ point decays into two acoustic phonons from the same branch while conserving both energy and momentum~\cite{klemens1966anharmonic}. Therefore, the acoustic phonon modes have $\hbar\omega_{0}/2$ energy with equal and opposite momentum. This process gives rise to a temperature dependent Raman shift of the following form: $\Delta \omega(T)\sim [1+\frac{2}{e^{\hbar\omega_{0}/2k_{B}T} - 1}]$~\cite{menendez1984temperature,balkanski1983anharmonic}. For both the $\mathrm{E^{1}_{2g}}$ and $\mathrm{A_{1g}}$ modes, with $\hbar\omega_{0}\approx 50$ meV. Thus, at low $T$ with $\Delta \omega$ saturates to a constant value, and becomes linear in $T$ at higher temperature. 

\begin{table*}[]
	\begin{tabular}{|l|c|c|c|c|c|c|c|c|}
		\hline
		& \multicolumn{4}{l|}{Experimental values($ cm^{-1}K^{-1} $)} & \multicolumn{4}{l|}{Theoretical values($ cm^{-1}K^{-1} $)} \\ \hline
		& $\alpha_{\mathrm{E_{2g}^1}} $    & $\alpha_{\mathrm{A_{1g}}} $   & $\beta_{\mathrm{E_{2g}^1}} $     & $\beta_{\mathrm{A_{1g}}} $& $\alpha_{\mathrm{E_{2g}^1}} $    & $\alpha_{\mathrm{A_{1g}}} $   & $\beta_{\mathrm{E_{2g}^1}} $     & $\beta_{\mathrm{A_{1g}}}$ \\ \hline
		SL on SiO$ _{2} $ 	   &0.0125     &	0.0133 &	0.0020    &	0.0083& 0.0041    &0.004	 &	0.0020    &	0.005  \\ \hline
		SL on hBN      & 0.0119     &	0.012 &  $\sim$0	       &	0.0049 &---     &	--- &  ---	       &	--- \\ \hline
		NBL on SiO$ _{2} $      &0.0123     &	0.0127 &	0.0028    &	0.0175 &0.0037     &0.0041	 &  0.0011	       &0.0024	  \\ \hline
		%ABL on SiO$ _{2} $     &0.006    &	0.0066 &	---  &	0.0048& ---     &	--- &  ---	       &	---  \\ \hline
	\end{tabular}
	{\caption{Temperature coefficient of Raman shift ($ \alpha $) and FWHM ($ \beta $) for SL and NBL on SiO$ _{2} $ for the two Raman modes of MoS$_2$.  
			\label{tab:1}}}
\end{table*} 

Note however, that the absolute values of the calculated mode frequencies differ from the experimentally measured ones by about 4\%. The frequency values are in fact sensitive to the choice of the exchange-correlation functional. Generalized gradient application (GGA) overestimates the lattice constant. For example, with GGA at 0~K the in-plane lattice constant is 3.18 {\AA}, which is $\sim 0.95\% $ larger than the experimentally measured value~\cite{wakabayashi1975lattice}, and the frequencies of the $E'$ and $A_1$ bands of SL are 373.7~cm$^{-1}$ and 396.2~cm$^{-1}$ while the experimental values at 8~K are 385.5~cm$^{-1}$ and 406.5~cm$^{-1}$.

Linear fits to the  calculated $T$ dependence of Raman shifts yield a temperature coefficient $ \alpha $  $\mathrm{\sim 0.004\ cm^{-1}/K}$ for both the $\mathrm{E_{2g}^1}$ and  $\mathrm{A_{1g}}$ modes (Table.~\ref{tab:1}). This value is significantly lower than the experimental values. This underestimation can arise from several parameters (both intrinsic and extrinsic), which are discussed below. 

Among the intrinsic phonon-phonon anharmonic effects ignored in our calculations, the principal one is a four-phonon processes. It has been proposed that, the relative contribution to Raman shifts at finite $T$ from the three-phonon and four-phonon processes is related to the phonon band-gap of the material. The larger the band-gap, the greater the contribution from four-phonon processes~\cite{Tianli_prbrap_2017}. The band-gap between acoustic and optical modes in SL $\mathrm{MoS_{2}}$ is $\sim 6$ meV, which is rather low. Thus, we expect that the phase-space available for three-phonon decay channels is non-negligible compared to higher order four-phonon processes. A significant contribution to the Raman shift temperature dependence from four-phonon and higher-phonon processes could thus be expected. Although extrinsic effects linked to defects~\cite{parkin2016raman} or the substrate~\cite{buscema2014effect}, adsorbates and fabrication induced disorder~\cite{mignuzzi2015effect} in a sample can give a constant shift to the Raman shift and FWHM of the Raman modes, we show in the next section that contribution to their temperature dependence is marginal.   

Compared to single layer, natural bilayer shows a red-shift in the $\mathrm{E^{1}_{2g}}$ Raman shift and a blue-shift in the $\mathrm{A_{1g}}$ Raman shift~\cite{doi:10.1021/nn1003937}. The blue-shift in the $\mathrm{A_{1g}}$ mode can be accounted for an additional spring-like interaction, related to short-ranged interaction involving the nearest neighbour S atom. On the other hand, the red-shift in the $\mathrm{E_{2g}^1}$ can be attributed to a greater dielectric screening of the Coulomb forces in $\mathrm{NBL \ MoS_{2}}$~\cite{molina2011phonons} compared to the case of single layer. The trends of the calculated temperature dependence of Raman shifts for NBL $\mathrm{MoS_{2}}$ look very similar to those of SL $\mathrm{MoS_{2}}$ (Figs.~\ref{Fig:4}(c) and (d)), in agreement with our experiment. 

\subsection{Effect of substrate on doping and strain levels in  MoS$_2$}	
	\begin{figure}[t]
	\begin{center}
		\includegraphics[width=0.45\textwidth]{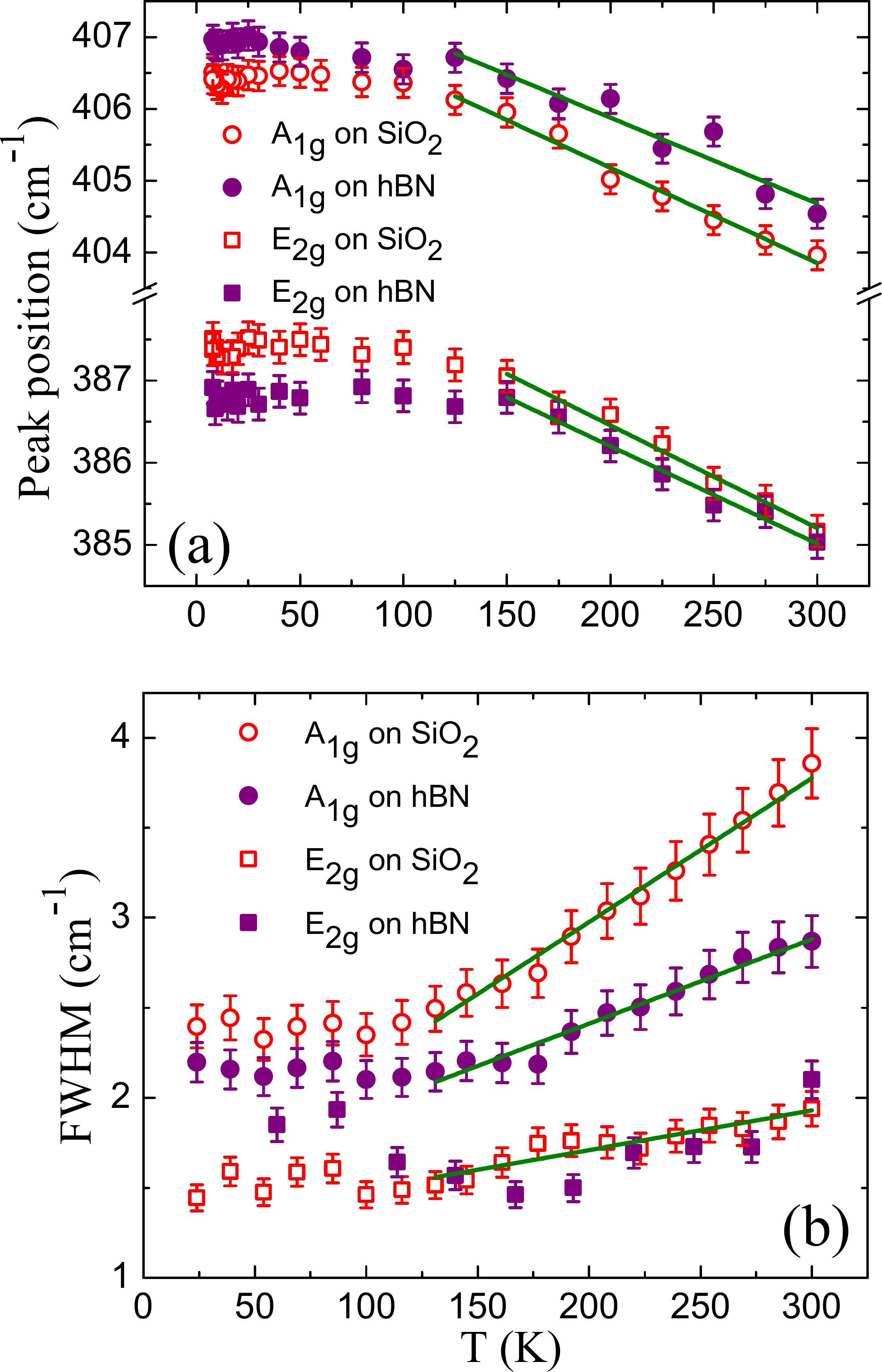}
		\small{{\caption{Plots of (a) the peak positions and (b) the FWHM of the $\mathrm{A_{1g}}$ and $\mathrm{E_{2g}^1}$ Raman modes of MoS$_2$ on hBN and SiO$_2$ substrates -- the data were taken on Sample2.  The green solid lines are the linear fits to experimental data for $T>100$~K. \label{Fig:5}}}}
	\end{center}
\end{figure}

The data discussed till now were obtained on MoS$_2$ on a SiO$_2$ substrate. We now turn to probe the effect of substrate on the Raman-modes. We have identified the single layer MoS$_2$ by comparing Raman spectra of SL MoS$_2$ on a SiO$_2$ with MoS$_2$ on hBN in Sample2 showed in the appendix section Fig.~\ref{Fig:12}. We performed temperature dependent Raman spectroscopy measurements in MoS$_2$ on hBN substrate -- the data are shown in Fig.~\ref{Fig:5}(a) (We have shown the Raman spectra for Sample2 measured at different temperature between 8~K to 300~K in appendix section Fig.~\ref{Fig:13}). The data show that the $\mathrm{E_{2g}^1}$ peak gets red-shifted for MoS$_2$ on the hBN substrate as compared to the SiO$_2$ substrate. On the contrary the $\mathrm{A_{1g}}$ peak gets blue-shifted for MoS$_2$ on the hBN substrate compared to the case with the SiO$_2$ substrate. We find a linear variation of $\mathrm{E_{2g}^1}$ and $\mathrm{A_{1g}}$ peak-frequencies with temperature above 100~K on the hBN substrate while at low-temperatures, the peak-frequencies and FWHM of both the  $\mathrm{E_{2g}^1}$ and $\mathrm{A_{1g}}$ saturate.  The fact that both the samples on SiO$ _{2} $ and on hBN show a saturation of Raman shift and FWHM at similar $T$-scale, points to an intrinsic origin. Finally we observe from Fig.~\ref{Fig:5}(b) that while  the FWHM of the  $\mathrm{E_{2g}^1}$ mode is comparable for Sample1 and Sample2, the FWHM of the  $\mathrm{A_{1g}}$ mode is significantly lower on the hBN substrate than on the SiO$_2$ substrate. In the following section we present our understanding of the origin of these observations.  

The induced carrier density and strain in $\mathrm{MoS_{2}}$ is substrate dependent. From our electrical transport measurements, we found that the mobility of SL MoS$_2$ on a SiO$_2$ substrate (on samples prepared in a way similar to Sample1) is $\mathrm{\sim2~cm^{2}V^{-1}s^{-1}}$ at 100 K while that of  SL MoS$_2$ on hBN (obtained on samples similar to Sample2) is $\mathrm{\sim20~cm^{2}V^{-1}s^{-1}}$~\cite{PhysRevB.99.245419}. The impurity number densities extracted from conductance fluctuation spectroscopy measurements for these two classes of samples are $\mathrm{3.5\times 10^{12}~cm^{-2}~eV^{-1}}$ and $\mathrm{1.8\times 10^{10}~cm^{-2}~eV^{-1}}$ respectively~\cite{PhysRevB.99.245419}. %Another report by Dubey et.al. ~\cite{doi:10.1021/acsnano.7b05520} have shown from their Raman shift analysis of A$_{1g}$ peak, the changes of electronic doping levels order of a few $\mathrm{10^{12} cm^{-2}}$ due to charge impurity on SiO$_2$ substrate.

The higher impurity concentration of  SL MoS$_2$ on a SiO$_2$ substrate effectively corresponds to a large electron doping. Consistently, the FWHM of the A$_{1g}$ peak is larger on the SiO$_2$ substrate than on the hBN substrate (Fig.~\ref{Fig:5}(b)).  In order to compute the effects of the electron doping on the Raman modes, we explicitly add a fraction of electron in the unit-cell of SL $\mathrm{MoS_{2}}$ in our calculations. We find that the $\mathrm{A_{1g}}$ mode softens significantly ( by $\sim -0.4 \ \mathrm{cm^{-1}}$ for 0.003e/cell) with electron doping in agreement with large electron-phonon coupling strength corresponding to this mode~\cite{chakraborty2012symmetry}. In sharp contrast, the $\mathrm{E^{1}_{2g}}$ mode is practically independent of doping (hardens by $\sim 0.08\ \mathrm{cm^{-1}}$ for 0.003e/cell). The temperature coefficient of the FWHM  of the Raman modes of hBN-MoS$_2$ sample actually compares well with the theoretically calculated slopes from three-phonon process. On the contrary, on a SiO$_2$ substrate, the experimental values are not well reproduced by our calculations, which yield significantly smaller estimates. In the case of hBN substrate charge impurity concentration is two orders of magnitude lower so electron-phonon processes are less pronounced. So the three-phonon process becomes the primary life-time determining mechanism in case of hBN-substrate device.

On the other hand, the strain induced due to $\mathrm{hBN}$ on MoS$_2$ is large as compared to that by $\mathrm{SiO_{2}}$. The lattice constant difference is more than 20 percent between hBN and $\mathrm{MoS_{2}}$. As the heterostructures are made by van der Waals interaction, we do not expect the strain is generated by lattice mismatch. Rather we believe the strain arises from deformed heterostructure due to fabrication. In our case we have prepared the heterostructure by stacking $\mathrm{MoS_{2}}$ on pre-transferred hBN flake by dry PDMS transfer technique. In this process the air been trapped in the interface of the heterostructure. As we anneal our sample through heating in vacuum at about 300$^\circ$C, the $\mathrm{MoS_{2}}$ got stretched by cleaning up most of the interface area and deformed the hBN. After annealing the $\mathrm{MoS_{2}}$ tries to relax but it got held by the deformed hBN. This does not allow the $\mathrm{MoS_{2}}$ to relax completely. As a result some residual strain sustain on the $\mathrm{MoS_{2}}$. Xu Han et.al. show  that in SL $\mathrm{MoS_{2}}$ the strain can be up to $0.6 \% $ in presence of hBN environment~\cite{han2019effects}. To investigate the effects of uniaxial compressive strain on the high-frequency Raman modes, we apply $0.1 \% $ to $0.4 \% $ strain to the unit-cell of SL $\mathrm{MoS_{2}}$ and compute the phonon mode frequencies using DFPT. We find that for MoS$_2$ on hBN substrate, the $\mathrm{E^{1}_{2g}}$ mode softens significantly ($\sim -1.3 \ \mathrm{cm^{-1}}$) more than the $\mathrm{A_{1g}}$ ($\sim -0.4 \ \mathrm{cm^{-1}}$). $\mathrm{SiO_{2}}$ as a substrate induces less strain, which results in a  stiffer $\mathrm{E^{1}_{2g}}$ mode compared to that of hBN. 
The combined effect of these two phenomenon -- namely higher electron doping levels on SiO$_2$ substrate and higher strain on hBN substrate --  provide a natural explanation of the experimentally observed red-shift for the $\mathrm{E_{2g}^1}$ peak and blue-shift in $\mathrm{A_{1g}}$ peak for MoS$_2$ on the hBN substrate as compared to that on the SiO$_2$ substrate (see Fig.~\ref{Fig:5}(a)). Interestingly, irrespective of the substrate, the temperature dependent Raman shift shows similar slopes overt the temperature range 100~K to 300~K signifying that the doping and strain does not play an important role in determining  the temperature dependence of Raman shift (Table.~\ref{tab:1}). This strongly supports our explanation of this trend using only intrinsic anharmonic processes.

\section{Conclusion}
We have performed a detailed study of temperature on Raman active modes in $\mathrm{MoS_{2}}$ and analyzed the effects of strain and electronic doping imposed by the substrate. Both the Raman shift and FWHM shows linear temperature dependence $T>100$~K, below $T<100$~K they became independent of the temperature. Using  first principle based calculations, we show that the observed temperature dependence of the Raman shift on SL and on NBL $\mathrm{MoS_{2}}$ arises from both three-phonon and four-phonon processes while the life-time of these phonon modes primarily arises from three-phonon process for the in-plane mode while for the out-of plane mode electron-phonon process plays an important role too. The higher order phonon processes are present in the system but life-time of those higher-order process is much longer than the three-phonon process due to much lower scattering probabilities and momentum and energy conservation rules. The higher value of FWHM in the out-of-plane vibrational mode as compared to that of the in-plane vibration is consistent with a scenario with a three-phonon process. To understand the contribution of other extrinsic effects like the presence of impurities and strain due to substrate on Raman shifts and Raman modes, lifetime a comparison of the data on samples fabricated on a hBN and on a $\mathrm{SiO_{2}}$ substrates has been performed. The theoretical calculations suggest that the observed differences arise from a larger strain and lower density of impurities on a hBN substrate. Lastly irrespective of sample quality, strain the temperature coefficient of Raman shift for in-plane and out-of-plane component is constant and it is linear throughout the temperature range 300~K to about 125~K. Below 100~K the Raman shift became independent of temperature and saturates. This extrinsic effect plays a static role in Raman-shift throughout the temperature range.

	\section{Acknowledgment}
	A.B. acknowledges financial support from SERB, DST, Govt. of India and Indo-French Centre for the Promotion of Advanced Research (CEFIPRA) and supports under FIST program, DST. The authors thank Supercomputer Education and Research Centre at IISc for providing computational resources.
	
	\section*{Appendix} 
	
		\subsection*{ Full range Raman spectra and identification of layer number on SiO$_2$ substrate }
		
		\begin{figure}[h!]
			\begin{center}
				\includegraphics[width=0.4\textwidth]{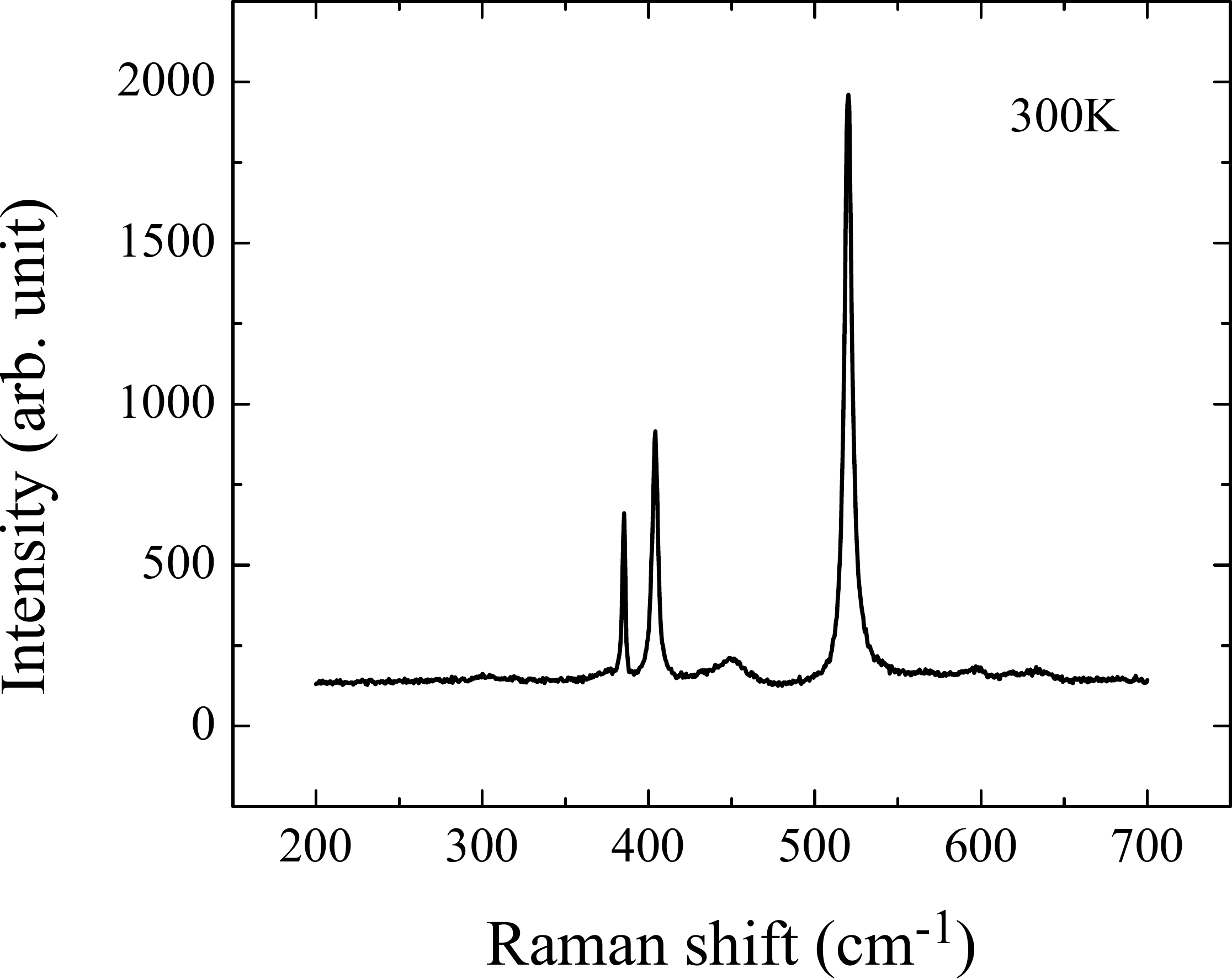}
				\small{{\caption { Raman spectrum for SL MoS$_2$ on SiO$_2$ substrate at 300~K.\label{Fig:6}}}}
			\end{center}
		\end{figure}
		
		In our main article, we have concentrated on the in-plane E$_{2g}$ and out-of-plane A$_{1g}$ vibrational modes. We have analyzed the Raman spectra only in the range of 370~cm$^{-1}$ to 420~cm$^{-1}$. But our measurement consist of the spectra ranges 200~cm$^{-1}$ to 800~cm$^{-1}$ for all the temperatures. In the Fig.~\ref{Fig:6} we have presented a full range spectra for SL MoS$_2$ on SiO$_2$ substrate at 300~K.

\begin{figure}[h!]
	\begin{center}
		\includegraphics[width=0.4\textwidth]{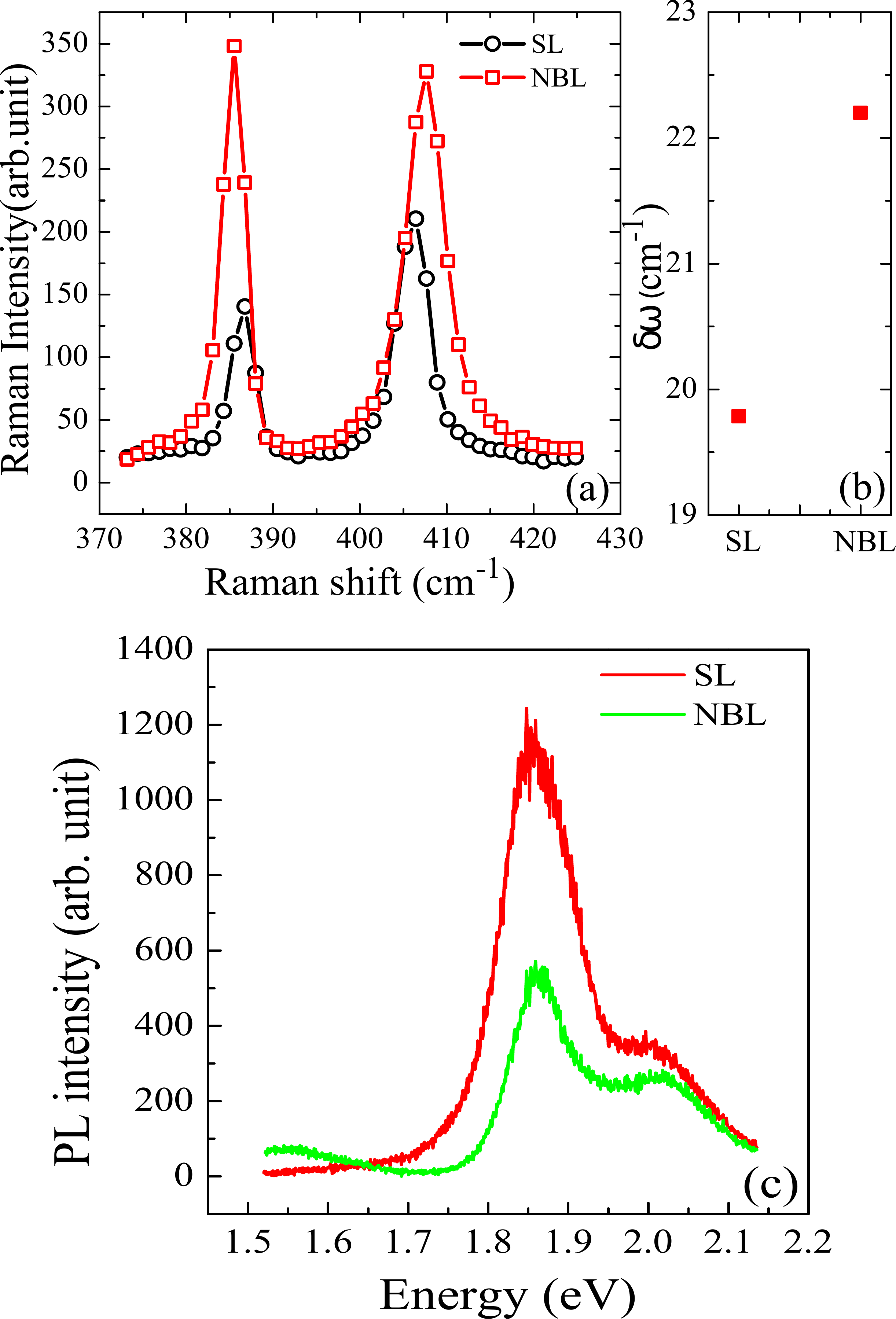}
		\small{{\caption {(a) Raman spectrum for SL and NBL MoS$_2$ on SiO$_2$ substrate at room temperature. (b) Raman shift $\delta\omega$ measured at SL and NBL MoS$_2$ on SiO$_2$ substrate. (c) Photoluminescence spectra for SL and NBL MoS$_2$ on SiO$_2$ substrate at room temperature.   \label{Fig:7}}}}
	\end{center}
\end{figure}	
	
	We have identified a single layer (SL) and natural bilayer (NBL) MoS$_2$ sample through Raman and PL spectra. In Fig.~\ref{Fig:7}(a) Raman spectra have been shown for SL and NBL samples. Raman spectra is fitted with two Lorentzian to identify the peak position and to calculate the difference in Raman shift between two peaks. We found that the difference in Raman shifts are 19.5 and 22~cm$^{-1}$ respectively shown in Fig.~\ref{Fig:7}(b) which are comparable to the previous report from the literature ~\cite{doi:10.1021/acsnano.7b05520}. On the other hand, the comparative PL spectra identified the SL and NBL samples.  SL has a much higher PL intensity than the NBL one shown in Fig.~\ref{Fig:7}(c). 
	
	\begin{figure}[tbh!]
		\begin{center}
			\includegraphics[width=0.4\textwidth]{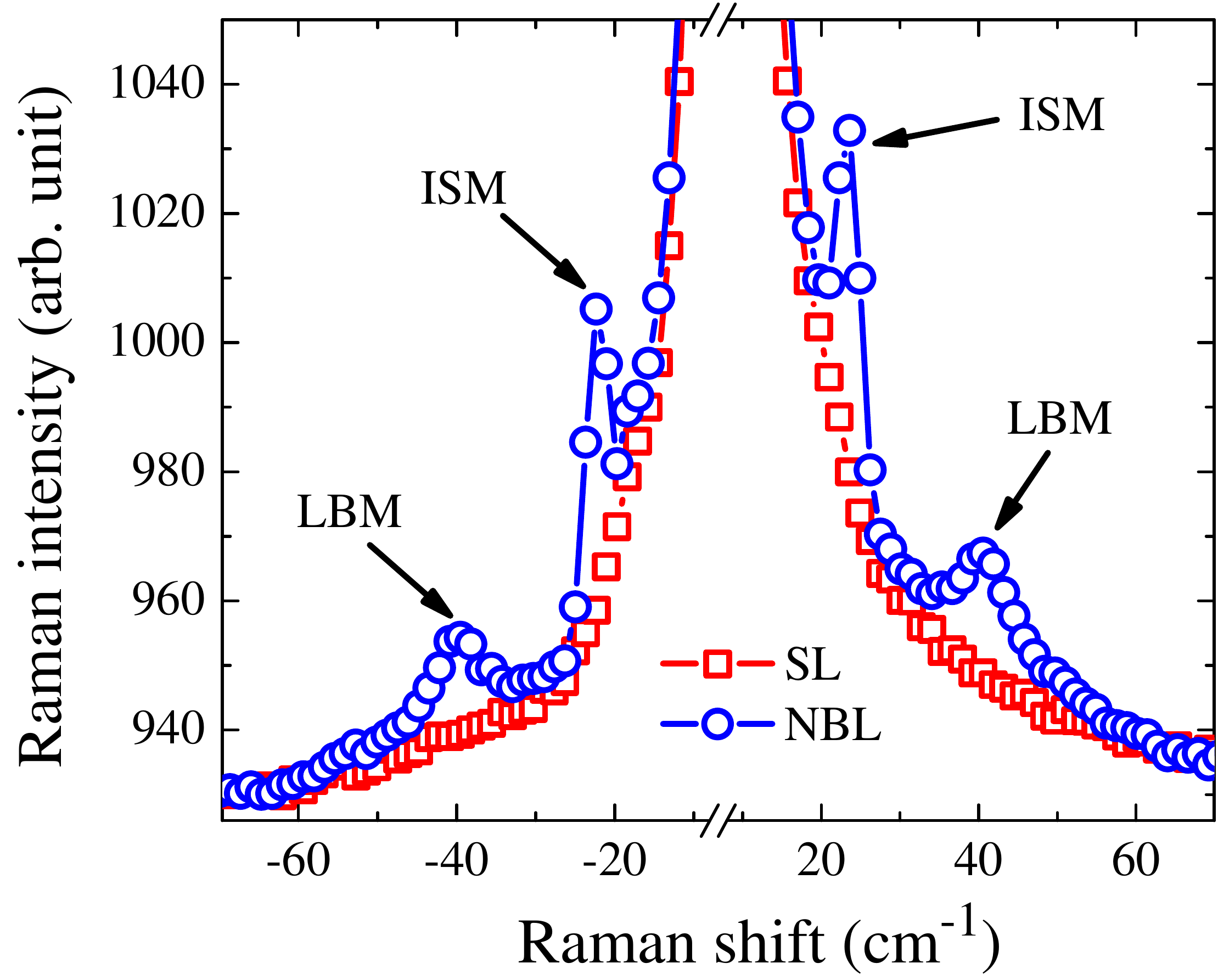}
			\small{{\caption { Raman spectra at room temperature in low frequency region for SL and NBL MoS$_2$ on SiO$_2$. We have marked the peak position of interlayer shear mode and layer breathing mode.  \label{Fig:8}}}}
		\end{center}
	\end{figure}
	
	We also have identified SL and NBL with low-frequency Raman measurement. In the case of SL sample, interlayer shear mode (ISM) and layer breathing mode (LBM) are absent while on bilayer sample ISM and LBM are present at about 20~cm$^{-1}$ and 40~cm$^{-1}$ respectively.The presence of interlayer modes confirms the layer number of the sample shown in Fig.~\ref{Fig:8}.

	\subsection*{ Results  obtained from Voigt fitting to the spectrum }
	
\begin{figure}[tbh!]
		\begin{center}
			\includegraphics[width=0.4\textwidth]{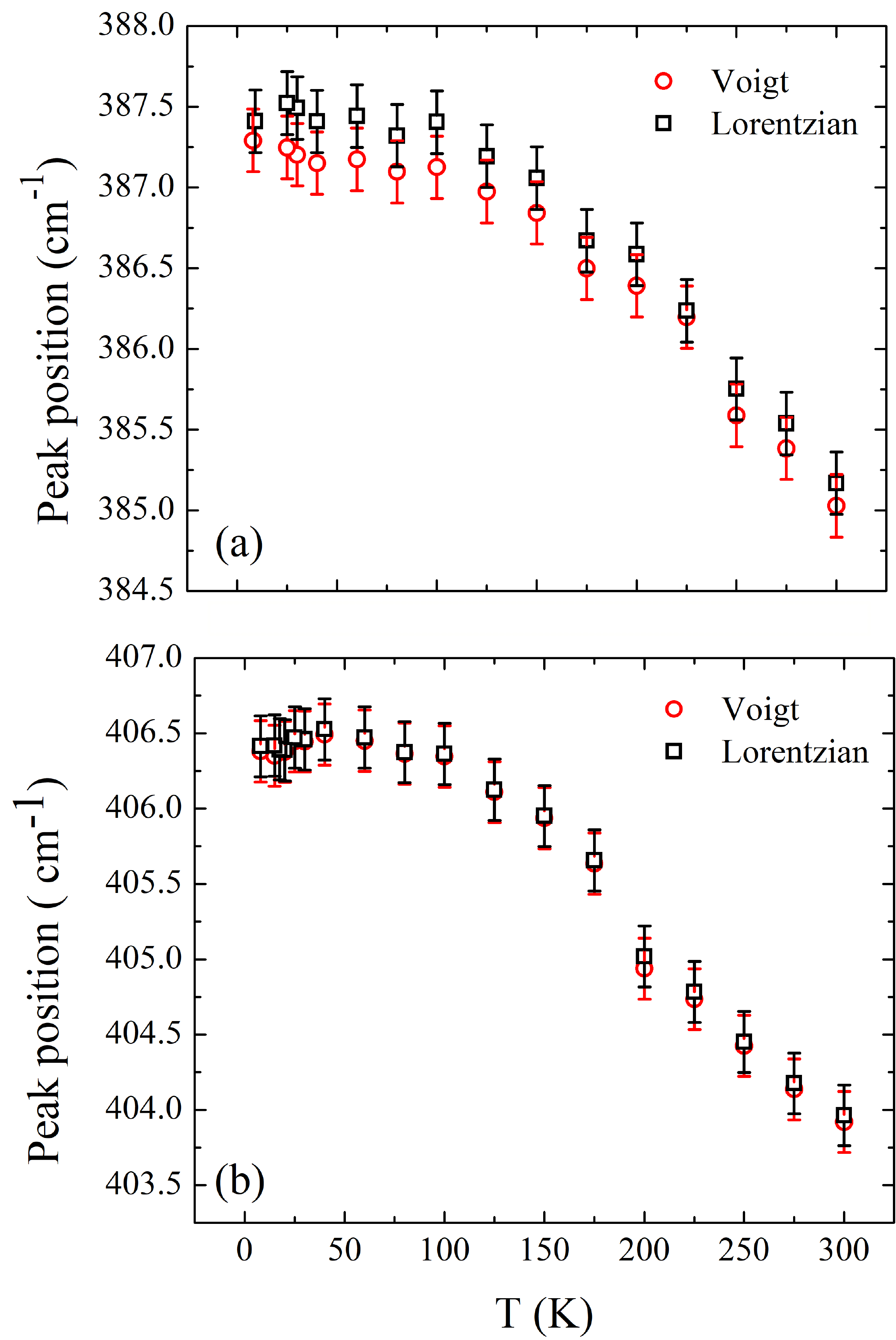}
			\small{{\caption { Plot of peak position with temperature for SL MoS$_2$ on SiO$_2$ by using Voigt and Lorentzian fitting to the Raman spectra for (a) E$_{2g}$ (b) A$_{1g}$ mode.   \label{Fig:9}}}}
		\end{center}
	\end{figure}

\begin{figure}[tbh!]
		\begin{center}
			\includegraphics[width=0.4\textwidth]{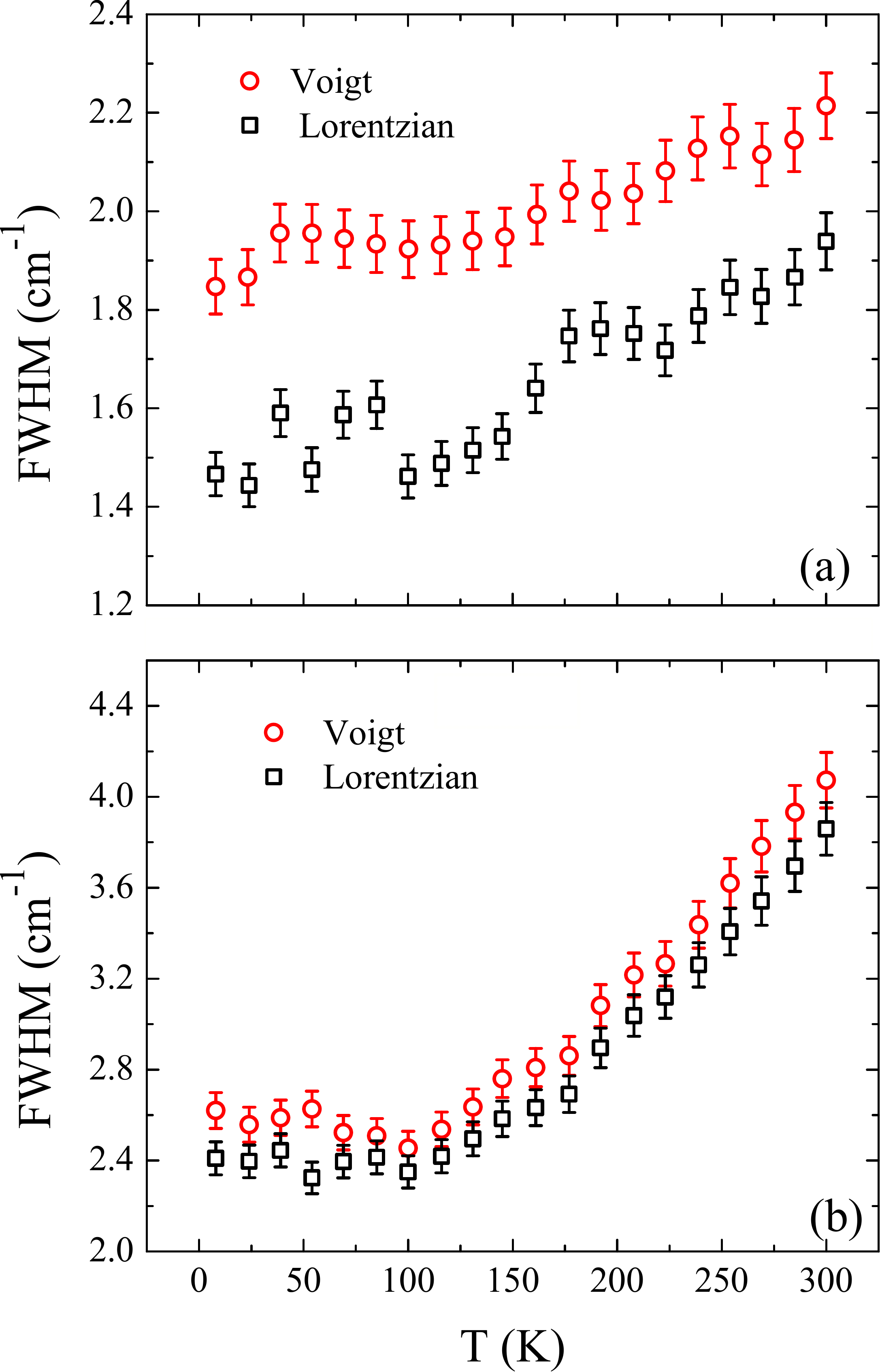}
			\small{{\caption { Plot of FWHM with temperature for SL MoS$_2$ on SiO$_2$ by using Voigt and Lorentzian fitting to the Raman spectra for (a) E$_{2g}$ (b) A$_{1g}$ mode.   \label{Fig:10}}}}
		\end{center}
	\end{figure}

	We have fitted our experimental result with Lorentzian in our manuscript. We have also used Voigt fitting to determine the peak position and FWHM. Peak position with temperature obtained from Voigt and Lorentzian fits are very similar for both E$_{2g}$ and A$_{1g}$ peak shown in Fig.~\ref{Fig:9}(a) and (b) respectively.
	We found that in case of Voigt fitting both Lorentzian width and Gaussian width contribute to the Voigt width and as a result, we end up with a minimum FWHM for E$_{2g}$ and A$_{1g}$ peak is about 1.8 cm$^{-1}$ and 2.5 cm$^{-1}$ respectively while from Lorentzian fit to our Raman spectra provides the minimum FWHM is about 1.45 cm$^{-1}$ and 2.3 cm$^{-1}$ respectively shown in Fig.~\ref{Fig:10}(a) and (b) respectively. In both Lorentzian and Voigt fitting FWHM follows the same trends with the temperature.
	
\begin{figure}[tbh!]
		\begin{center}
			\includegraphics[width=0.4\textwidth]{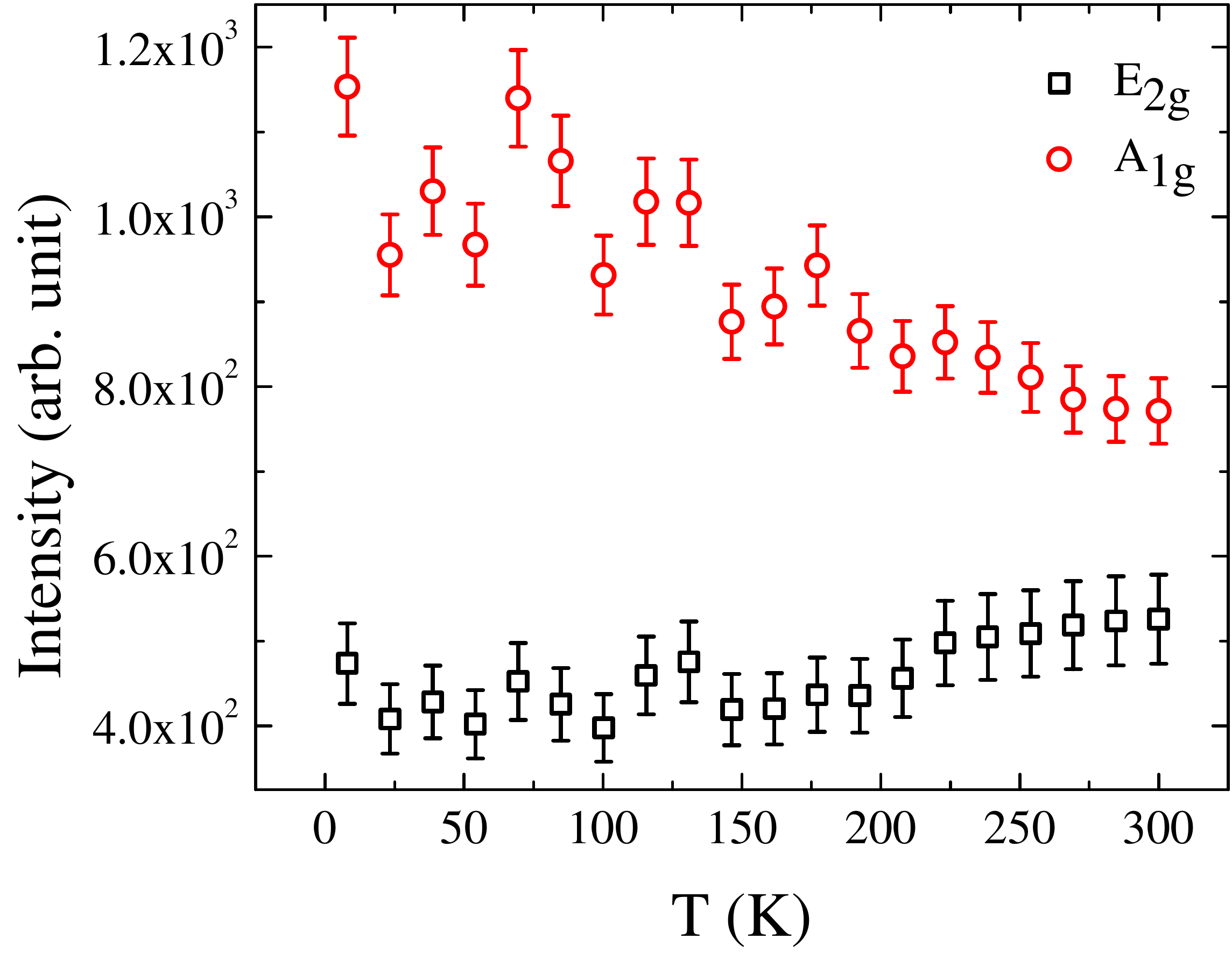}
			\small{{\caption { Plot of peak intensity with temperature for E$_{2g}$ and A$_{1g}$ mode in SL MoS$_2$ on SiO$_2$ substrate.  \label{Fig:11}}}}
		\end{center}
	\end{figure}

	We have plotted the intensity of the Raman peaks for E$_{2g}$ and A$_{1g}$ mode for SL  MoS$_2$ on SiO$_2$ substrate with temperature in  Fig.~\ref{Fig:11}. The peak intensity for E$_{2g}$ peak remains almost unchanged with temperature while A$_{1g}$ peak intensity decreases as we increase the temperature. 
	
	\subsection*{ Detection of SL MoS$_2$ on hBN substrate and temperature-dependent Raman spectra  }
\begin{figure}[tbh!]
		\begin{center}
			\includegraphics[width=0.4\textwidth]{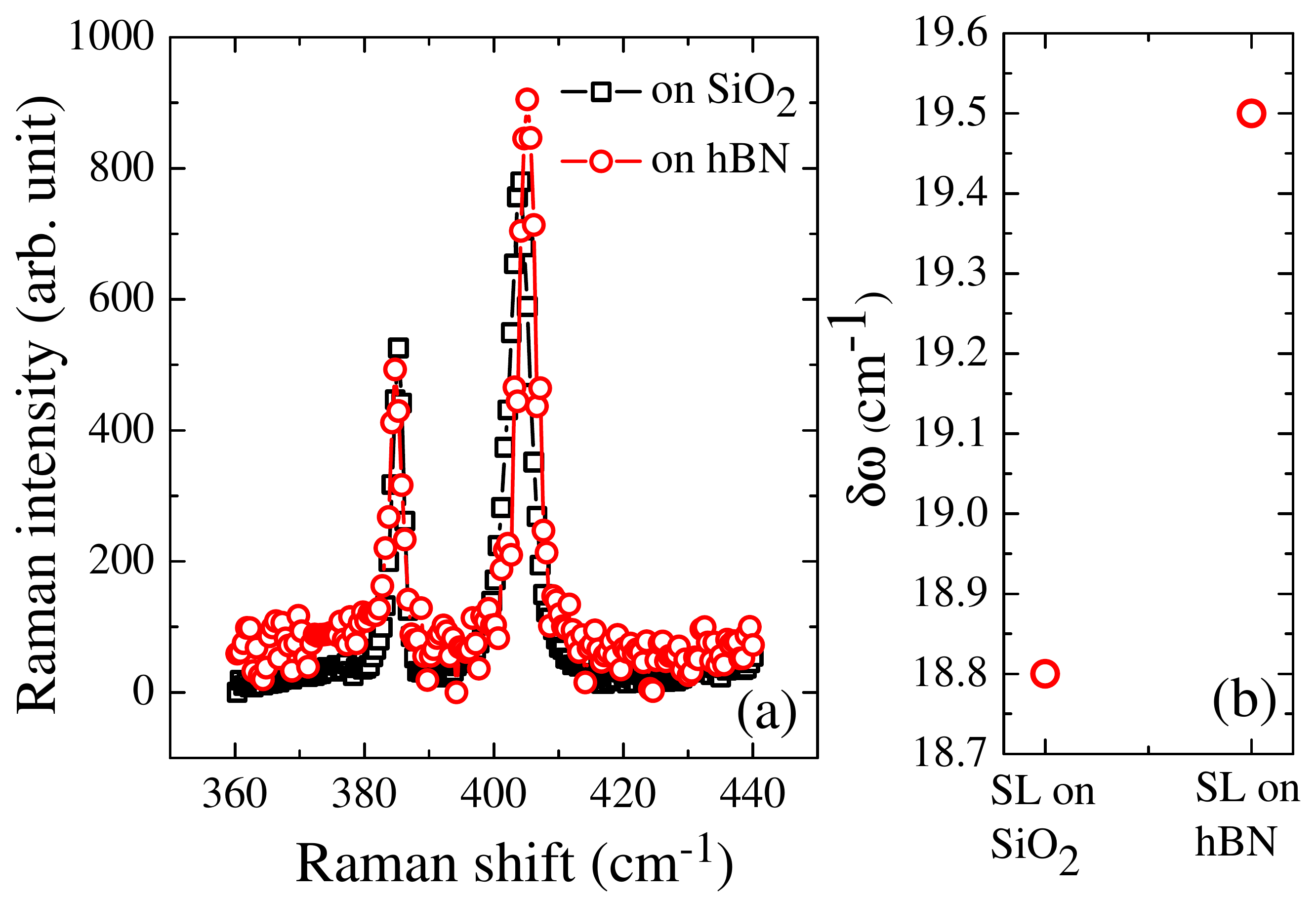}
			\small{{\caption { Plot of (a) High frequency Raman spectrum of SL MoS$_2$ on SiO$_2$ and hBN substrate. (b) High frequency Raman shift $\delta\omega$ measured at SL MoS$_2$ on SiO$_2$ and hBN substrate. \label{Fig:12}}}}
		\end{center}
	\end{figure}

	We have identified SL MoS$_2$ on the hBN substrate and compared the Raman spectra with the SiO$_2$ substrate sample shown in Fig.~\ref{Fig:12}(a) and calculated the peak position difference. We found that the peak position difference is 18.8~cm$^{-1}$ and 19.5~cm$^{-1}$ for SiO$_2$ and hBN substrate sample respectively shown in Fig.~\ref{Fig:12}(b).

\begin{figure}[tbh!]
		\begin{center}
			\includegraphics[width=0.5\textwidth]{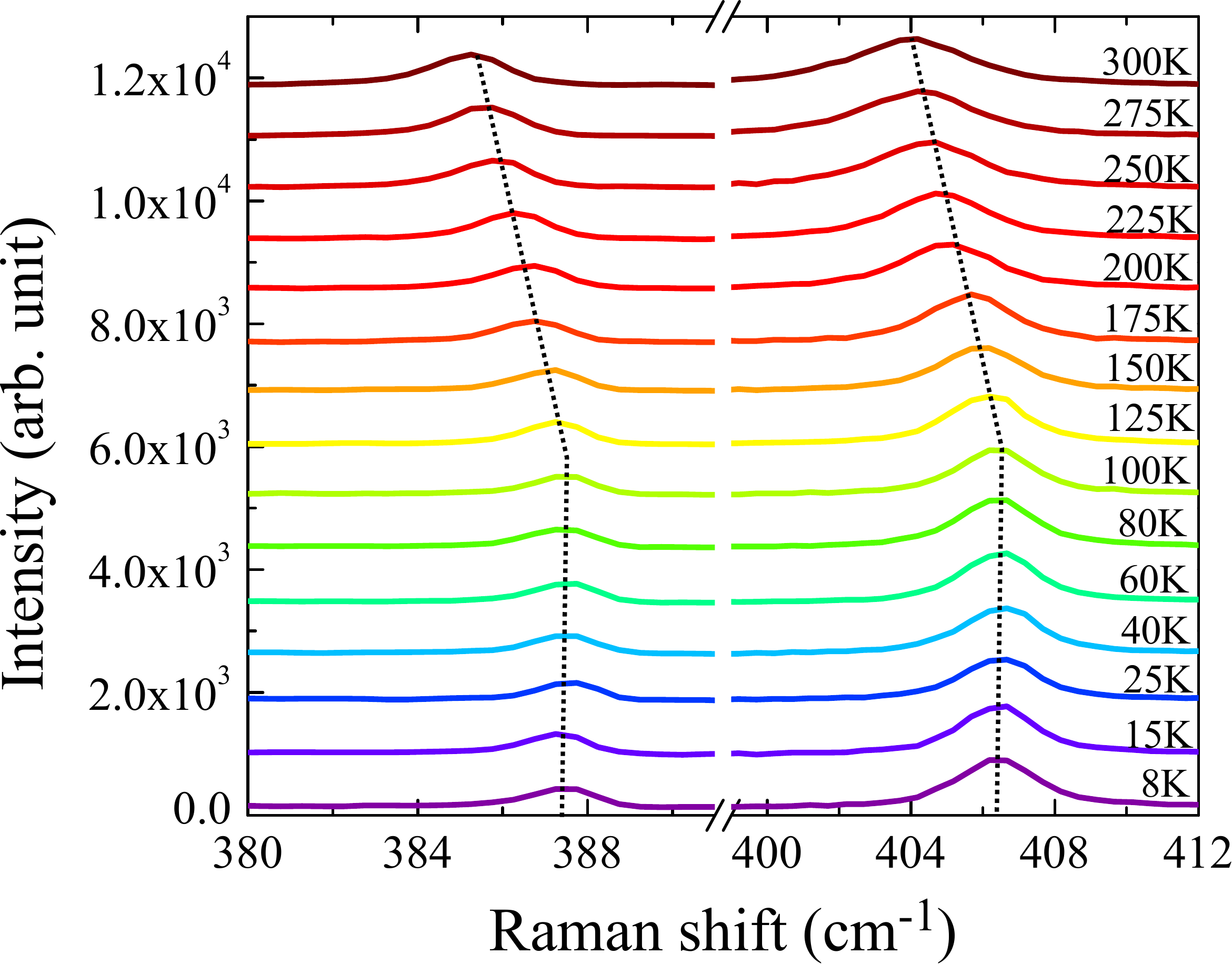}
			\small{{\caption {Raman spectra of hBN substrate SL MoS$_2$ on Sample2 measured at different temperatures between 8~K and 300~K. The dotted lines are the guide for the eye for marking the evolution of Raman shift with temperature. \label{Fig:13}}}}
		\end{center}
	\end{figure}

	We have presented the temperature dependent Raman spectra for hBN substrate SL MoS$_2$ on sample2 from 300~K to 8~K temperature range shown in Fig.~\ref{Fig:13}. The spectra shows the blue shift with temperature till 100~K and then saturates below that temperature. The FWHM decreases with decreasing temperature can also be noticed from the data.

\end{document}